\documentclass{aa}

\usepackage{graphicx}
\usepackage{txfonts}
\usepackage{placeins}

\linenumbers

\begin{document}

   \title{Extreme value distribution for gamma-ray-burst prompt data}

   \subtitle{How unexpected was the BOAT event?}

   \author{S. Covino
          \inst{1,2}
          }

   \institute{INAF - Osservatorio Astronomico di Brera, via E. Bianchi 46, 23807 Merate, Italy\\
              \email{stefano.covino@inaf.it}
            \and
             Como Lake centre for AstroPhysics (CLAP), DiSAT, Università dell’Insubria, via Valleggio 11, 22100 Como, Italy
             }

   \date{Received September 15, 1996; accepted March 16, 1997}

 
  \abstract
   {Gamma-ray bursts (GRBs) are known to be unpredictable in time and position. A few (observationally) exceptional events have been observed, such as \object{GRB\,221009A,} which stands out for having a fluence and peak flux orders of magnitude higher than what has been measured so far.}
   {Analysing the observed fluence, peak flux, or duration distributions typically requires one to assume some scenarios, and the consistency of the observed data with the predictions turns out to be an important model diagnostic. However, it is also of interest to model these distributions using general statistical properties that do not rely on specific model assumptions, allowing one to derive inferences only based on the consistency of the observed distributions with the hypothesis of one single population of events that generate them.}
   {We obtained fluences, peak fluxes, and durations from the catalogues of GRBs observed by the CGRO-BATSE and Fermi-GBM instruments. We selected the extreme values in slots of equal duration and modelled their distributions using the generalised extreme value (GEV) formalism. The GEV distribution is a limit distribution naturally arising when the number of observations is large and is essentially independent of the phenomena producing the observed data.}
   {The distributions of extreme values for fluences, peak fluxes, and durations are consistent with being extracted from a single population of events, but the fluence and peak flux recorded for \object{GRB\,221009A} constitute a striking exception. The probability of observing such an event, assuming it is a cosmological GRB, is low, with a median value of about one event per millennium for the fluence and about one event per century for the peak flux.}
   {}


   \maketitle
%

\section{Introduction}

Gamma-ray bursts (GRBs) have been routinely observed by high-energy instruments since their discovery
by the Vela early warning satellites \citep{Klebesadeletal1973}. Since then, a fairly large set of space-borne instruments have monitored the sky at times, generating catalogues with thousands of events, such as those reported by the Burst and Transient Source Experiment \citep[CGRO-BATSE,][]{Goldsteinetal2013},\footnote{\url{https://heasarc.gsfc.nasa.gov/W3Browse/cgro/bat5bgrbsp.html}}  Konus-Wind \citep[e.g.][]{Tsvetkovaetal2017}, and
the Fermi Gamma-Burst Monitor \citep[Fermi-GBM, ][]{Gruberetal2014,vonKienlinetal2014,Bhatetal2016,vonKienlinetal2020,Poolakkiletal2021}.\footnote{\url{https://heasarc.gsfc.nasa.gov/W3Browse/fermi/fermigbrst.html}}

Gamma-ray bursts show two usually distinct emission phases: the prompt and the afterglow \citep{Zhang2018}. They have been classified in (at least) two different groups, including the long-soft and the short-hard GRBs \citep{Kouveliotouetal1993,Qinetal2013}, originating from two different classes of progenitors \citep{Zhang2018}. However, exceptions to a classification based only on duration and hardness ratios have been singled out \citep[see e.g.][]{Ahumadaetal2021,Zhangetal2021,Yangetal2022,Dichiaraetal2023,Levanetal2024,Yangetal2024,Yietal2025}.

The availability of such a large collection of data has naturally allowed for a large variety of statistical analysis concerning, for example, their sky distribution \citep{Briggsetal1996}, their duration and hardness \citep{Kouveliotouetal1993}, and their luminosity function \citep{Fenimoreetal1993,Salvaterraetal2012} or relations between prompt and/or afterglow parameters \citep{Yonetokuetal2004,Amati2006,Ghirlandaetal2012,Dainottietal2023}. In this framework, the rather recent detection of \object{GRB\,221009}, with its extreme properties in terms of fluence and peak flux \citep[see e.g. ][]{Frederiksetal2023,Burnsetal2023,Lesageetal2023}, represents a formidable test bench for any attempt to model the observed population(s) of cosmological GRBs, and even the possibility of physics beyond the standard model \citep{Galantietal2024,Galantietal2025}.

This exceptional GRB \citep[e.g. ][and references therein]{Malesanietal2023} was rapidly dubbed the Brightest Of All Times (BOAT). Indeed, it was extremely bright GRB, producing a fluence and peak flux (after correction for saturation) more than an order of magnitude greater than any other event detected so far \citep[][and references therein]{Burnsetal2023}. This event is of course interesting by itself, but several studies have also been devoted to establishing the probability of observing such an event if \object{GRB\,221009A} were a regular cosmological GRB. The results unavoidably depend on the various assumptions involved in the analysis. In \citet{Malesanietal2023}, rates range from one case every few decades up to a century. Instead, \citet{Burnsetal2023} obtained, by geometric extrapolation of the $\log(N)-\log(S)$ distributions for a multi-mission catalogue of long GRBs, that the fluence of \object{GRB\,221009A} makes it a once-in-ten-thousand-years event. Estimates in 1 for $10^2$ to $10^4$ years have been obtained by several other authors \citep[e.g. ][]{Atteiaetal2025,Lanetal2023,O'Connoretal2023,Williamsetal2023}. Such a low rate stimulated the researchers to find different ways to avoid the (statistical) contradiction between the actual observation and the length of the sky monitoring (a few decades). \citet{Bloom2022} suggested that \object{GRB\,221009A} could be the result of Galactic lensing. \citet{Finke&Razzaque2024} discussed the possibility that \object{GRB\,221009A} belonged to a different population of narrow-beam GRBs, so that its occurrence rate could be as high as approximately one per 200\,yr. Eventually, the possibility that this event was not a cosmological GRB has also been discussed \citep[e.g., ][]{Naviaetal2024}.

The problem of determining the true rate of events comparable to \object{GRB\,221009A} therefore remains, and it comes with many consequences for our understanding of the cosmological GRB population. In this paper, we propose a different, almost model-independent approach; i.e. based only on the assumptions that the detection probability of bright GRBs by high-energy satellites is essentially constant during the lifetime of the mission and that the probability of such an event is also constant in time (i.e. it is stationary). In fact, we modelled the set of bright events observed by the CGRO-BATSE and Fermi-GBM catalogues using the so-called generalised extreme value distribution \citep[GEV; ][]{Beirlantetal2004,Coles2001}; i.e. a statistical distribution naturally emerging asymptotically and almost independently of the phenomena under study. In Sect.\,\ref{sec:gev}, we describe the properties of the GEV distribution. In Sect.\,\ref{sec:dt}, we introduce the data set used in this work, and we discuss the analysis carried
out Sect.\,\ref{sec:an}. Finally, in Sect.\,\ref{sec:disc}, we comment our results and present our conclusions.


\section{The GEV distribution}
\label{sec:gev}

The extreme value theory, i.e. the attempt to evaluate the probability of occurrence of extreme measurements, is a rapidly evolving field of modern statistics. It is applied to many scenarios such as climatology, engineering, physics, and economy \citep{Coles2001}. In astronomy, recent applications include the modelling of the brightest galaxies in a survey \citep{Heatheretal2024}, the number of sunspots \citep{Zhangetal2024}, the maxima of periodograms in time-series analysis \citep{Gondhalekaretal2023,Suveges2014}, density fields in cosmology \citep{Repp&Szapudi2018}, masses of galaxy clusters \citep{Waizmannetal2012}, and so on.

The most interesting aspect of the GEV distribution is that it can be proved that it holds asymptotically, essentially independently of the probability distribution describing the phenomena that generate the observed data. Obtaining mathematical proof is somewhat tedious \citep[e.g.][]{Beirlantetal2004}, but not difficult. There is some analogy with the validity of the central limit theorem, i.e. that under rather general conditions the limiting distribution of sample means for a large family of parental statistical distributions is a standard normal distribution \citep{Fischer2011}.

If we collect the maxima, $Z_{\mathrm{max},n}$, of a large number of random variables, $Z_1, \ldots, Z_n$, identically distributed according to a given continuous distribution, $F(z)$, with $n \longrightarrow +\infty$, the GeV distribution can be written as
\begin{eqnarray} \label{eq:gev}
G(z) &= & {\rm Pr} \{ Z_{\max,n} \leq z  \}  \nonumber \\
         &= &\exp  \left\{ - \left( 1 + \xi \frac{z - \mu}{\phi} \right)^{-1/\xi} \right\}, \nonumber \\
  &   & \xi \in \mathbb{R}, \quad \mu \in \mathbb{R}, \quad \phi > 0,  \nonumber
\end{eqnarray}
where $1+\xi(z-\mu)/\phi > 0$.

The parameter $\xi$ is usually called {\it \emph{shape,}} and it is related to the tail decay of the underlying distribution of $F(z)$. $\mu$ is the {\it \emph{mean}} value of the distribution and $\phi$ is the {\it \emph{scale}}. The {\it \emph{shape}} divides the GEV family into three sub-families. Negative {\it \emph{shape}} parameters signal distributions of maxima of variables with a finite upper boundary (i.e., $\mu - \phi/\xi)$, while a positive or zero {\it \emph{shape}} parameter indicates that there is no upper limit. The case $\xi = 0$ is also known as the `Gumbel' distribution. It is defined as the limit function when $\xi \rightarrow 0$, and it takes the form
\[
G(z) = \exp\left\{ - \exp \left( - \frac{z-\mu}{\phi} \right) \right\},
\]
with $z \in \mathbb{R}$.

A GEV model is typically trained by dividing a dataset into blocks of uniform length (e.g. a year, etc.). Then, from each block of data, the extreme (maximum or minimum) is extracted, and finally a GEV model is fitted to the set of extreme values. Once trained, the GEV distribution can be used to calculate return levels. The return level associated with the return period, $1/p$, is simply the level expected to be exceeded once in all $1/p$ blocks of the same length $k$. Credible intervals for any estimated quantity can be derived by the posterior distribution of the GEV parameters.

The choice of the length of the blocks used to extract extreme values is not driven by first principles, and there is a well-known bias-variance trade-off. The GEV is a limit distribution; therefore, longer blocks --i.e. including a large number of elements-- provide a better approximation. On the other hand, longer blocks, for any finite set of observed data, imply a reduced number of elements and unavoidably a poorer prediction with a larger variance. Smaller blocks might, on the contrary, give a worse approximation of a GEV distribution, yielding biased predictions. Therefore, the choice of block length, unless driven by physical reasons, requires a trial-and-improvement process; the results are tested using appropriate model diagnostics such as, in most cases, the quantile-quantile (QQ) plot \citep[e.g.][]{Coles2001}. If the model is good, the points will closely follow a straight line with intersect 0 and slope 1 without strong systematic deviations.

We mention, finally, that a crucial assumption of extreme value analysis is that the original distribution of data that extreme values are extracted from should be stationary. If this does not happen, it is indeed possible to include exogenous variables in the analysis to model the non-stationarity values \citep[e.g. trends, seasonality, etc.; ][]{Coles2001}.

\section{Data set selection}
\label{sec:dt}

In this work, we based our analyses on the Fermi-GBM GRB catalogue \citep[e.g.][]{Poolakkiletal2021}, and we considered the CGRO-BATSE GRB catalogue \citep{Goldsteinetal2013} as a comparison case with no events comparable to \object{GRB\,221009A}. The GBM catalogue contains 4029 events from July 2008 to June 2025,\footnote{Updated up to June 26 2025.}  while the BATSE catalogue contains 2145 events from April 1991 to August 1996.
From these catalogues, we extracted the trigger time, the fluence, the peak flux, and the GRB duration; the latter is expressed, as usual, by the parameter $T_{90}$, the duration of the time interval during which 90\% of the total observed counts were detected. In more detail, from the BATSE catalogue we considered the total fluence in the energy range 20-300\,keV (summing the fluences recorded in the 20-50, 50-100, and 100-300\,keV bands), in units of erg cm$^{-2}$ and the peak flux in the 20-300\,keV energy range on the 1024\,ms timescale in units of photons cm$^{-2}$ sec$^{-1}$. From the GBM catalogue we considered  the fluence in the 10-1000\,keV energy range, in erg cm$^{-2}$ and the peak flux in the same energy range on the 1024\,ms timescale in units of photons cm$^{-2}$ sec$^{-1}$. Choosing different baselines for the peak flux computation does not change any of the conclusions reported in this paper.

Since we are interested in the extreme values for the considered parameters, it is necessary to discuss the various biases affecting data obtained from the selected catalogues. In fact, the brightest events, are often affected by saturation. A correction for the saturation is typically carried out comparing results from different instruments by a suitable cross-calibration, and/or a modelling of the spectral evolution of the event in order to infer the expected behaviour when the considered instrument was unable to provide reliable information. In most cases, such a correction is of limited entity and not important for the present study. \object{GRB\,221009A} is an exception due to its peculiar brightness. A saturation-corrected fluence was computed in \citet{Lesageetal2023}, and the reported value is more than two times higher than the value reported in the GBM catalogue. A saturation-corrected peak flux was computed by \citet{Frederiksetal2023} based on Konus-Wind data in an energy band close to the Fermi-GBM one. This estimate turns out to be an order of magnitude higher than what was reported in the GBM catalogue. In the following analysis we adopted the saturation-corrected values for the BOAT.

In addition, the BATSE catalogue is known to be incomplete towards the highest fluxes and fluences \citep[see e.g.][]{Kanekoetal2006,Burnsetal2023} due to several factors including saturation and triggering strategy. This unavoidably affects our capability to derive a solid GEV distribution for the parameter of our interest, and incompleteness in the derived extreme values can make the extrapolation towards the highest fluences and peak fluxes more uncertain beyond the statistical uncertainties. On the other hand, as far as the bulk of the high-fluence and -flux reported events are reliable, the global features of the computed GEV distributions should be still of interest.

Independently of the considered catalogue, events with very long duration might suffer from biases due to various possible constrains affecting instrument efficiency aboard of low orbit satellites (e.g. Earth occultation, South-Atlantic anomaly crossing, telemetry limitations, etc.). Catalogues are therefore likely truncated, missing, or incorrectly record some of the longest duration GRBs. The BATSE catalogue does not report events longer than $\sim 700$\,s, while the GBM catalogue seems to be limited at $\sim 1000\,s$, indicating the higher level of incompleteness suffered by the former. A few ultra-long-duration GRBs ($>> 1000\,s$) have also been singled out \citep[e.g. ][]{Levanet2014}, opening a long debate about their possible different origin compared to the bulk of cosmological GRBs. Deriving the true duration of a very long GRB can be a difficult task, often relying on data from multiple missions. Inclusion of these data in the present analysis could undoubtedly be of interest, but it would also require proper homogenisation since we should derive a duration seen by the given (BATSE or GBM) instrument without limitations; i.e. a different quantity with respect to the duration of emission activity of the source at any energy. Since the focus of the present work is on the possible peculiarity of \object{GRB\,221009A}, whose duration is well within the recorded values in the GBM catalogue, we decided not to apply any correction to the duration data.

\section{GEV analysis}
\label{sec:an}

The choice of the block length to extract the maxima, as discussed in Sect.\,\ref{sec:gev}, is not driven by first principles and is a trade-off between the goodness of the GEV distribution fit versus the variance of the prediction. The higher the number of events per block (i.e. the longer the block duration), the better the GEV distribution fit and the worse is the variance on the prediction. In our cases, ablock length of 30\,days seems to offer a good compromise given the average number of events per day in both catalogues ($\sim 0.6-0.7$\,day). For the BATSE catalogue, we obtained a sample of 66 extreme values. The richer GBM catalogue instead allowed us to derive a sample of 207 extreme values. The selected 'extreme' events are shown in Fig.\,\ref{fig:gev}.

For both the considered catalogues, the selected extreme values have a signal-to-noise ratio better than $\sim 10$, making their uncertainties irrelevant for the considerations derived in this work.

We performed a Bayesian analysis and adopted improper uniform priors for the model parameters. The resulting posterior is proper as long as the sample size is larger than 3 \citep{Northrop&Attalides2016}. Sampling is carried out by the `No-U-Turn Sampler' extension \citep{Hoffman&Gelman2014} to the Hamiltonian Monte Carlo algorithm \citep{Brooksetal2011}. The sampled chains in all cases converge to equilibrium, and the posterior distribution of parameters shows a peaked distribution for the three parameters of the model. Corner plots for the six cases considered in this work are shown in Fig.\,\ref{fig:corners}, and summary statistics for the GEV-distribution fitted parameters are reported in Table\,\ref{tab:fitres}.

In several cases, the {\it \emph{shape}} parameter distributions include the zero values and the simple G\"umbel distribution in their possible ranges. However, some of the distributions seem to be better fitted by negative {\it \emph{shape}} parameters, although in no case is the evidence very strong. This essentially implies that the available data sets are not able to unambiguously indicate whether their extreme value distributions are characterised by an asymptotic maximum value or are likely ever increasing going to an infinite number of blocks. There are also instrumental limitations to consider in evaluating the obtained results. For example, the difficulties in recording long uninterrupted observations makes the lack of very long duration GRBs in the considered catalogues, at least partly, artificial, likely pushing the {\it \emph{shape}} parameter to negative values.

Since the fluence and peak flux of \object{GRB\,221009A} are highly discrepant with respect to the population of GRBs from the GBM catalogue (see also Sect.\,\ref{sec:disc}), we also carried out the fit excluding the BOAT from the data set. This is, at this stage, a rather arbitrary choice, but, quite interestingly, without \object{GRB\,221009A} the GEV distribution for the fluence would indicate an upper boundary value at $\sim 0.08-0.09$\,erg\,cm$^{-2}$. This figure is close to the fluence recorded for the BOAT. The analogous limit for the peak flux is instead much higher than the recorded value, and, as expected, the GEV distribution for the duration is almost unaffected by the inclusion of \object{GRB\,221009A} in the analysis.

\begin{figure*}
\centering
\begin{tabular}{cc}
\includegraphics[width=0.7\columnwidth]{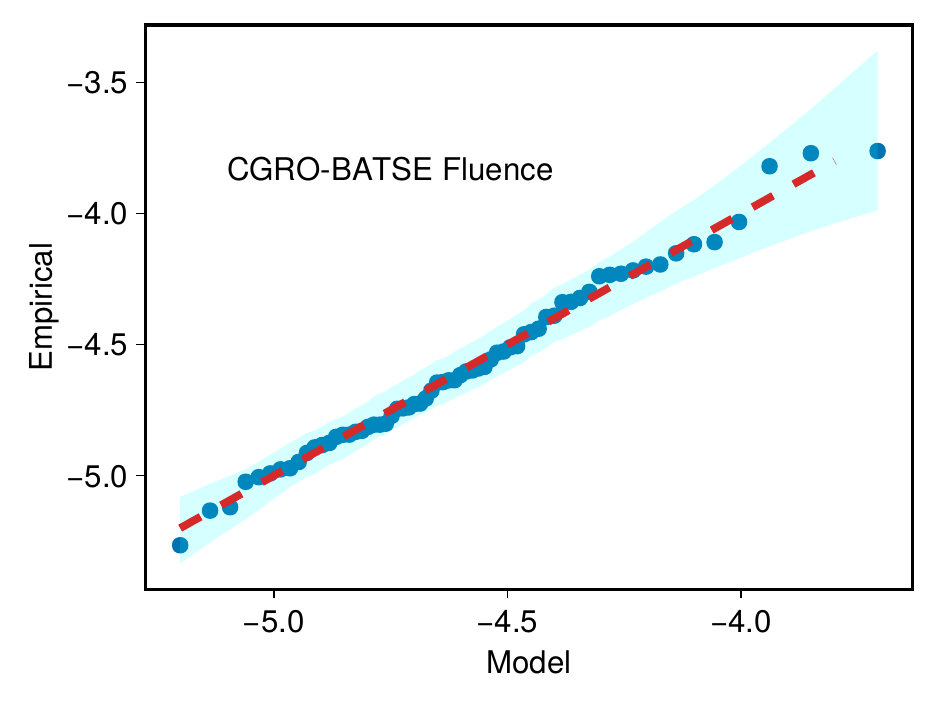} &
\includegraphics[width=0.7\columnwidth]{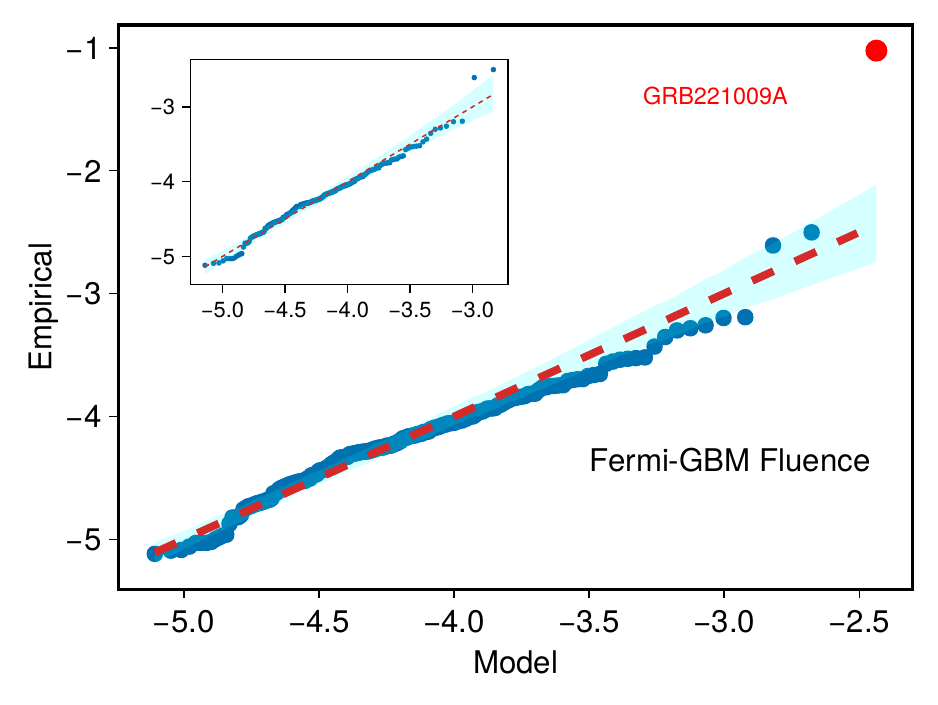} \\
\includegraphics[width=0.7\columnwidth]{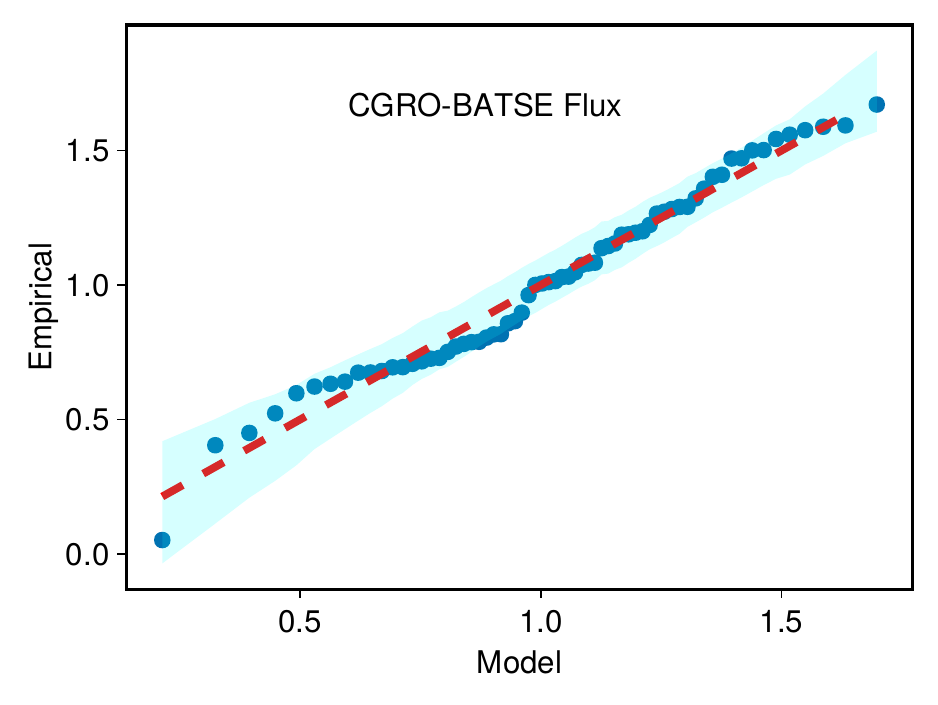} &
\includegraphics[width=0.7\columnwidth]{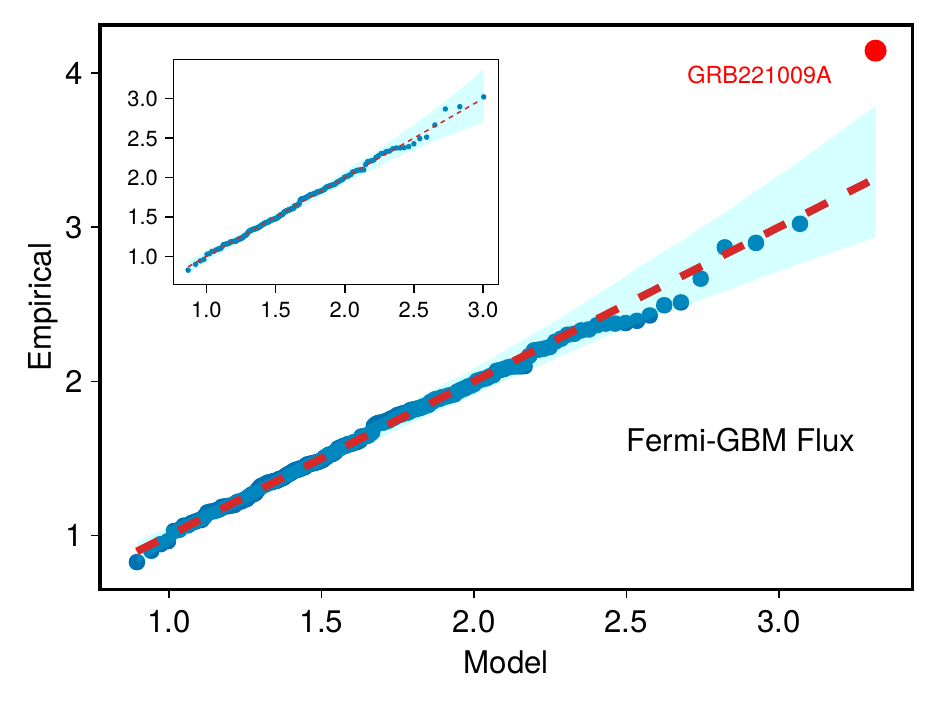} \\
\includegraphics[width=0.7\columnwidth]{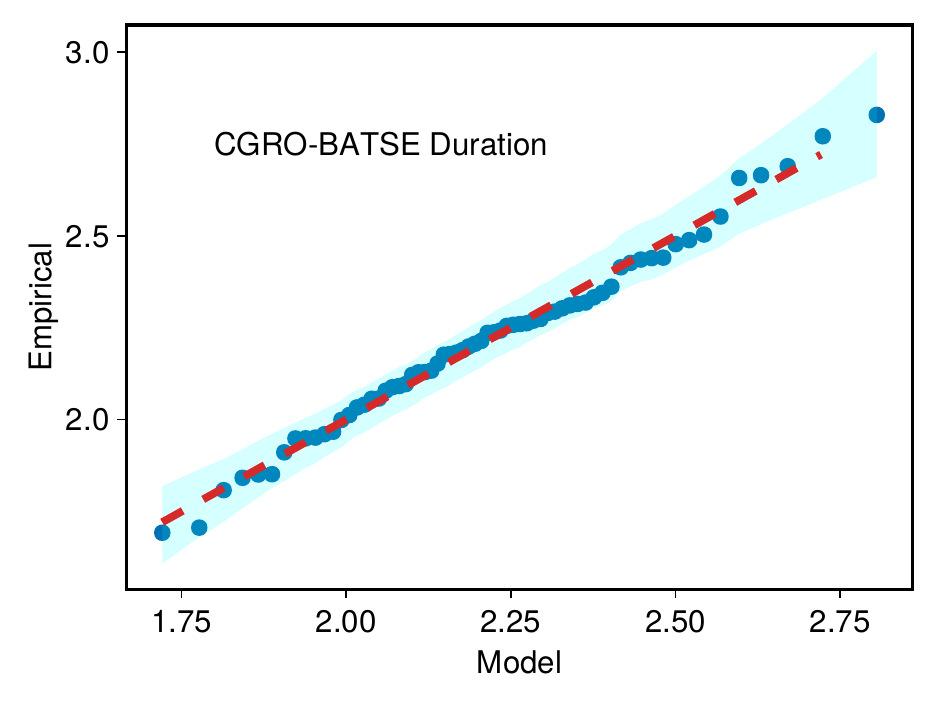} &
\includegraphics[width=0.7\columnwidth]{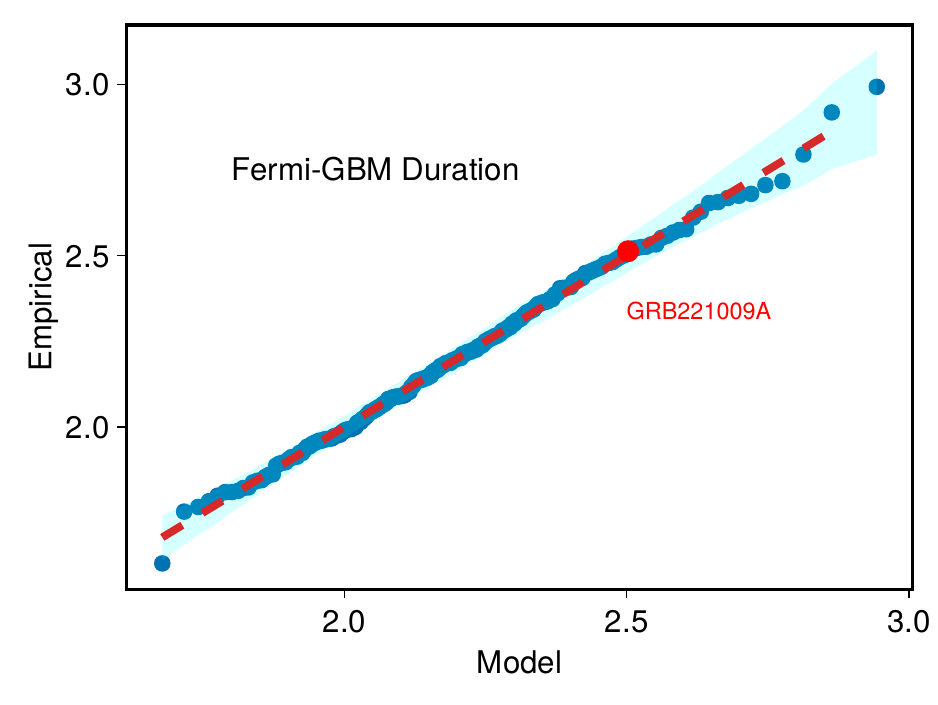}
\end{tabular}
\caption{QQ plots for GEV fit of extreme values extracted from the duration (bottom), peak flux (middle), and fluence (top) of the CGRO-BATSE catalogue (left) and Fermi-GBM catalogue (right). The 95\% credible region is also shown. \object{GRB\,221009A} (red symbol) is clearly an outlier for the fluence and peak flux, but it is not for the duration. The insets show the results of the fit excluding \object{GRB\,221009A} from the data set. \object{GRB\,230307A} and \object{GRB\,130427A}, respectively the second and third event with the highest fluence in the Fermi-GBM catalogue, are on the bright side, yet still consistent with GEV distribution (they are within the 99\% credible region for a data set of 207 events).}
\label{fig:qq}
\end{figure*}

As discussed in Sect.\,\ref{sec:gev}, the fit quality was checked by diagnostic plots, as were the widely known QQ plots. For both the data sets considered in this work, the distributions of durations, peak fluxes, and fluences turn out to be nicely fit by a GEV distribution (Fig.\,\ref{fig:qq}), with the already-mentioned notable exception of  \object{GRB\,220109A} for the fluence and the peak flux that clearly stands out among the extreme values from the GBM catalogue. As already pointed out in past analyses \citep[e.g.][]{Burnsetal2023,Malesanietal2023}, the peculiarity of \object{GRB\,220109A} is due to a combination of factors; for example, persistent high flux emission, long duration, and the spectral hardness.

\begin{figure*}
\centering
\begin{tabular}{cc}
\includegraphics[width=0.8\columnwidth]{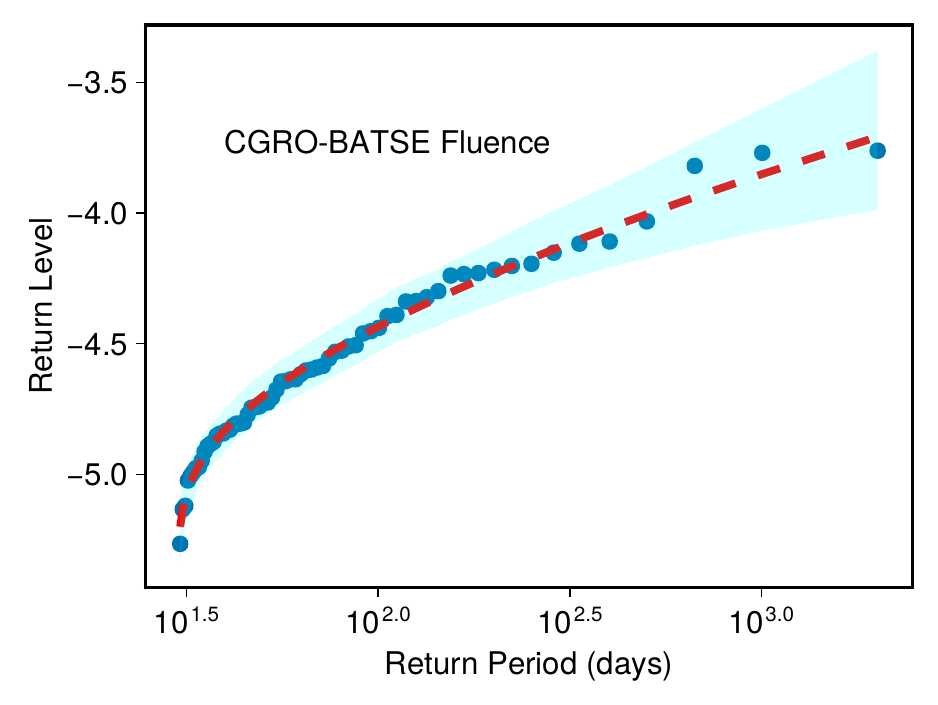} &
\includegraphics[width=0.8\columnwidth]{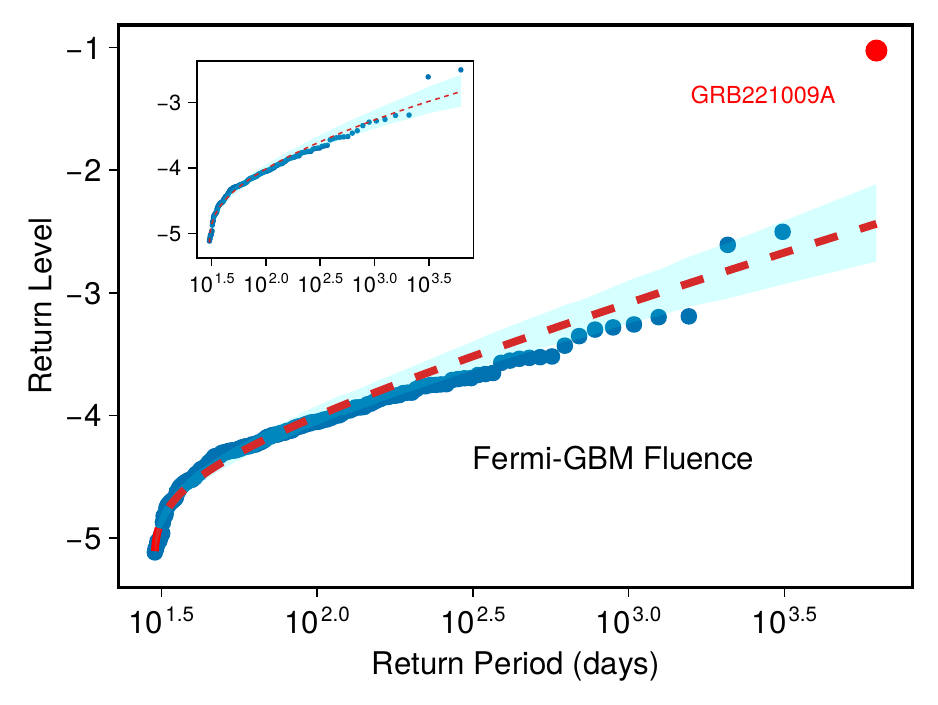} \\
\includegraphics[width=0.8\columnwidth]{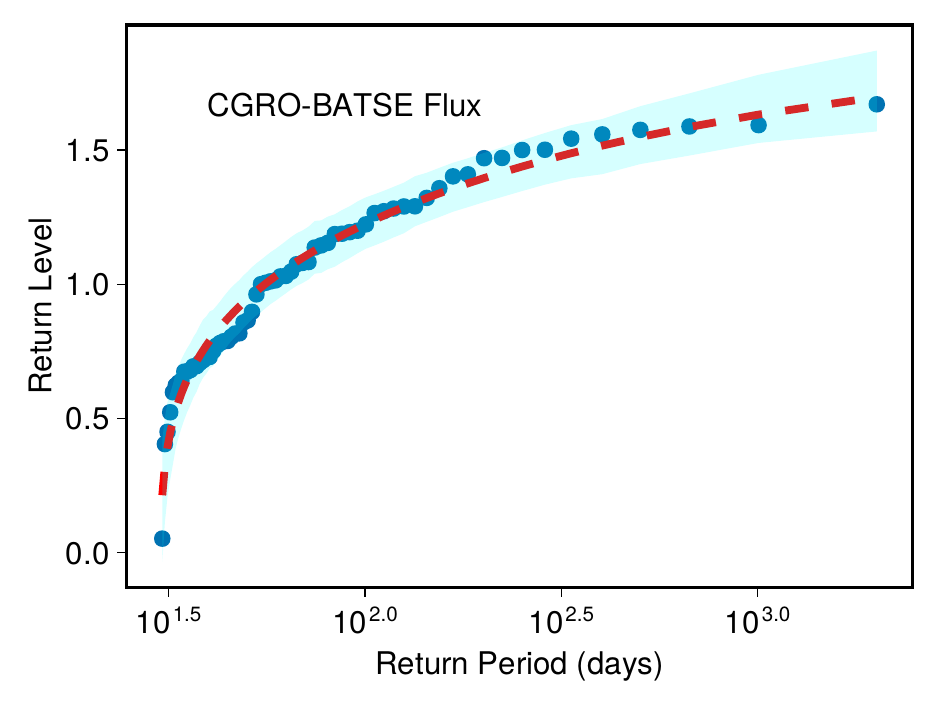} &
\includegraphics[width=0.8\columnwidth]{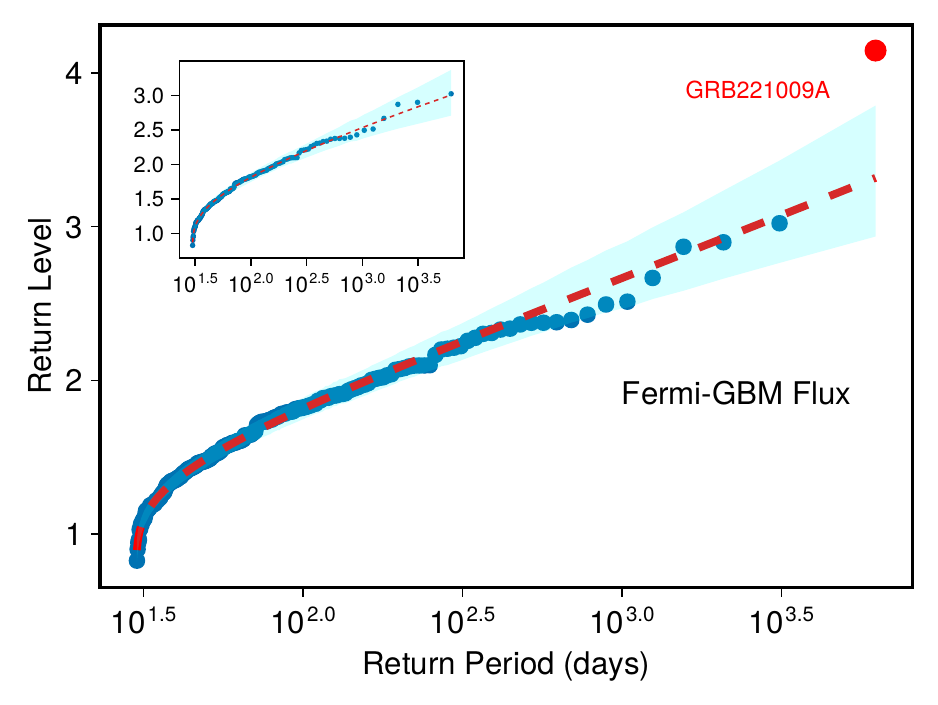} \\
\includegraphics[width=0.8\columnwidth]{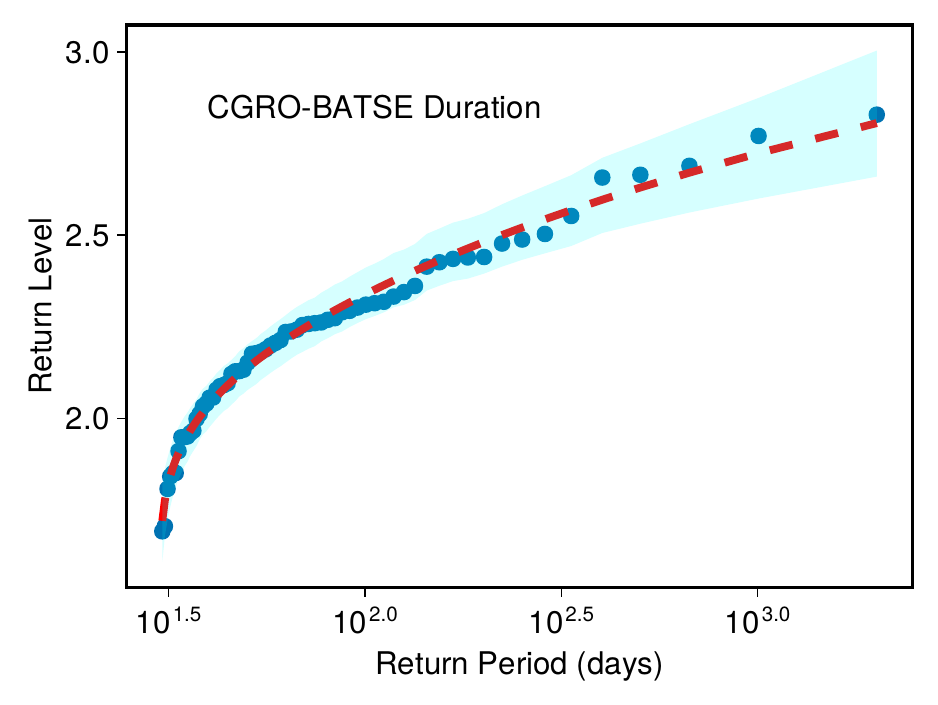} &
\includegraphics[width=0.8\columnwidth]{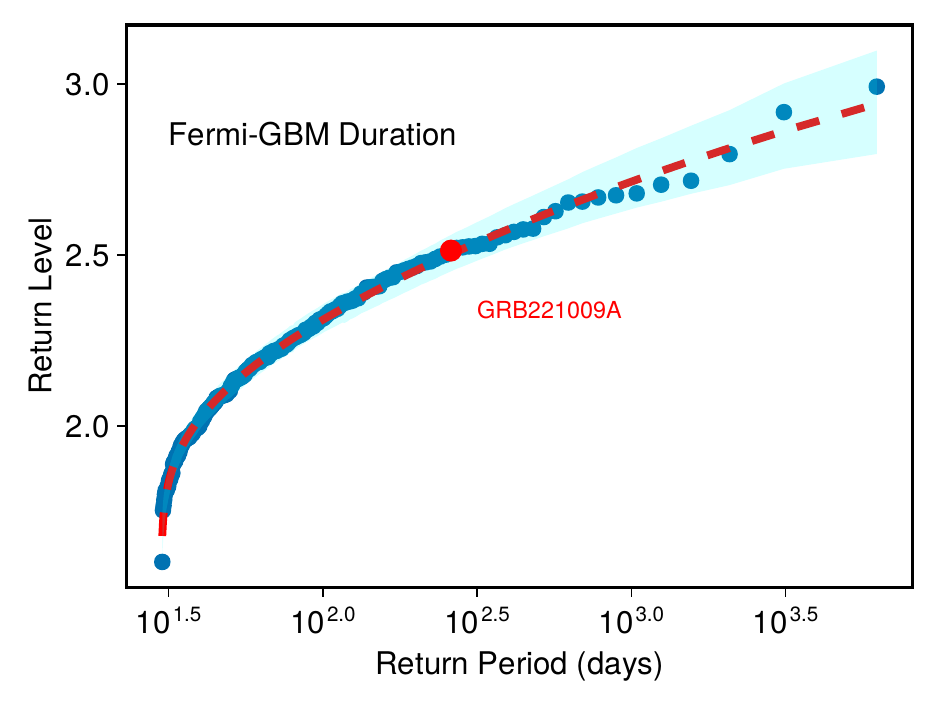}
\end{tabular}
\caption{Return plots for GEV fit of extreme values extracted from the duration (bottom), peak flux (middle) and fluence (top) of the GRO-BATSE catalogue (left), and Fermi-GBM catalogue (right). The 95\% credible region is also shown. The model cannot interpret the return level of \object{GRB\,221009A} for the fluence and peak flux while this event is part of the general distribution for duration. The insets show the results of the fit excluding \object{GRB\,221009A} from the data set. \object{GRB\,230307A} and \object{GRB\,130427A}, respectively the second and third event with the highest fluence in the Fermi-GBM catalogue, are on the bright side, yet they are still consistent with GEV distribution (they are within the 99\% credible region for a data set of 207 events).}
\label{fig:rt}
\end{figure*}

Finally, the return plots are shown in Fig.\ref{fig:rt}. The reported levels are expected to be exceeded on average once every $1/p$ years; i.e., the return level is exceeded in any particular block with a probability of $p$.

For the BATSE catalogue, as expected from the analysis of the QQ plot, the return levels are all well consistent with the modelling (Fig.\,\ref{fig:rt}). Things are considerably different for the GBM catalogue data, where \object{GRB\,220109A} again appears grossly inconsistent with the model.

Just extrapolating the fit distribution, the median value of the model return level, at the fluence of \object{GRB\,220109A}, is more than $\sim 1000$ years, and the BOAT fluence enters the 95\% allowed range at $\sim 150$\,years. A peak flux value comparable to what was shown by \object{GRB\,220109A} is expected in $\sim 140$\,years, while the observed peak flux enters the 95\% range in $\sim 30$\,years. Removing  \object{GRB\,220109A} the GEV distributions for the fluence and peak flux gives a negligible probability of observing an event of comparable fluence or peak flux for timescales of thousands years.

\section{Discussion and conclusions}
\label{sec:disc}

The most interesting feature of a GEV analysis is that it does not require any assumptions beyond that the analysed data are drawn from a stationary distribution. Durations, peak fluxes, and fluences for the brightest events detected by a specific instrument satisfy this requirement, since, to a large extent, we can assume that the natural degradation of the instruments have a minor or even negligible impact on their capability to detect observationally bright events.
In addition, focusing on the longest, brightest, and highest fluence events, we are also automatically excluding GRBs belonging to the short-duration class from the data sets \citep{Kouveliotouetal1993}, with the partial exception of the extreme values for the peak fluxes. Quite interestingly, for both the considered catalogues,  $\sim$14\% (BATSE) and $\sim8\%$ (GBM) of peak flux extreme values are due to events shorter than the canonical 2\,s value.  On the other hand, in spite of the fact that the brightest events are also, unavoidably, the best studied, there is still the (more or less virtual) possibility that the BATSE and GBM samples are polluted by very bright events due to phenomena erroneously classified as GRBs. As a matter of fact, it was also proposed that some of the brightest events come from a different (and sub-dominant) population of GRBs \citep[e.g.][]{Finke&Razzaque2024,Brieletal2025}. Analogous considerations have been proposed for the ultra-long GRBs ($\gg$ 1000\,s) \citep[e.g. ][]{Levanet2014}, which, however, are not included in the present analysis.
 
Our results indicate that, apart from the case of \object{GRB\,221009A}, which we discuss in the next paragraph, the GEV distributions for the GRB reported in the BATSE and GBM catalogues are consistent with being drawn from a single population of events. Their respective GEV distributions are also substantially consistent with each other (Table\,\ref{tab:fitres}). For instance, the return level for a GRB duration of $\sim 1000$\,s is close to one event in 20-30\,years in both cases, which is possibly inconsistent with the few ultra-long GRBs observed so far, although the inefficient recording of very-long-duration events makes any conclusions about this topic at this stage simply preliminary.

As already observed, the data for \object{GRB\,221009A} are well consistent with the population GEV distributions if we refer to the duration. They are instead highly discrepant for the fluence and peak flux. This is a non-trivial point, and it shows that the peculiarities of the BOAT are due to the combination of several factors such as long-lasting high-flux emission \citep[and its spectral hardness, e.g.][]{Frederiksetal2023} rather than to a very long $T_{90}$ or very bright flux only. This is also in agreement with findings from previous works \citep[e.g.][]{Malesanietal2023,Burnsetal2023}. The conclusions about the peak flux hold even if different baselines for the flux computations are used.

The fluence and peak flux of \object{GRB\,221009A} are, therefore, grossly inconsistent with the GEV distribution. The median of the return level for the fluence yields a probability of one event of this fluence occurring in about 1000\,years, while for the peak flux the median return value is about 140\,years. All in all, \object{GRB\,221009A} appears an unlikely event for its fluence and peak flux, given the duration of the monitoring of the considered missions. On the contrary, if we adopt the hypothesis that an event with fluence and peak flux as large as or larger than \object{GRB\,221009A} is unlikely, this might imply that one of the assumptions of the present analysis --i.e. that the measured extreme values are drawn from a homogenous population of GRBs-- is violated. As a matter of fact, the possibility that \object{GRB\,221009A} is not a typical cosmological GRB has been proposed and discussed in the literature. This might be due to a different geometry of the event \citep{Finke&Razzaque2024}, a misclassification of the event as a GRB \citep{Naviaetal2024}, or just low-probability intense gravitational-lens enhancement \citep{Bloom2022}. In all cases, these hypotheses require their rate of occurrence to be low enough not to affect the bulk of the GRB population observations and to mainly affect the fluence and, to a lesser extent the peak flux, but not the duration of the detected events.

An analysis considering all the GRBs detected by any instrument after a proper homogenisation is in principle possible, although the possibility of introducing artefacts due to incomplete knowledge of emission spectra and the intrinsic limitations of the involved instruments is high. We leave this possible development to a future work.

\section{Data availability}

We developed software tools and used third-party
libraries coded in the {\tt julia} language \citep[v. 1.13.4-1.13.5, ][]{Bezansonet2017}.\footnote{\url{https://julialang.org/}} GEV models and fitting tools are from the {\tt Extremes} package\footnote{\url{https://jojal5.github.io/Extremes.jl/dev/}} \citep[v. 1.0.3-1.0.5,][]{Jalbertetal2024}. Plots are produced by the {\tt makie}\footnote{\url{https://docs.makie.org/v0.22/}} \citep[v. 0.22.2-0.24.0,][]{Danisch&Krumbiegel2021} with the {\tt CairoMakie} backend, and corner plot are produces by the {\tt PairPlots}\footnote{\url{https://sefffal.github.io/PairPlots.jl/dev/}} \citep[v. 3.0.1,][]{Thompson2023} packages. Project management was carried out by the {\tt DrWatson}\footnote{\url{https://juliadynamics.github.io/DrWatson.jl/dev/}} package \citep[v. 2.18.0, ][]{Datserisetal2020}. Bibliographic research has also been aided by the {\tt Google Gemini} (v. 2.5) IA model.\footnote{\url{https://gemini.google.com/}}

Data and code used in this work are available at the gitlab repository: \url{https://www.ict.inaf.it/gitlab/stefano.covino/grb-extreme-values.git}.

\begin{acknowledgements}
      I thank dr. Giancarlo Ghirlanda and Om Sharan Salafia for useful discussions and dr. Eric Burns for kindly sharing his datasets. I also thank the anonymous referee. Her/his pertinent comments have substantially improved this manuscript. I also acknowledge the contribution by the Italian Space Agency, contract ASI/INAF n. I/004/11/6.
\end{acknowledgements}


\bibliographystyle{aa}
\bibliography{mybib}

\begin{appendix}

\section{Extreme values}
\label{sec:extreme}

\begin{figure*}[hp!]
\centering
\begin{tabular}{ccc}
\includegraphics[width=0.6\columnwidth]{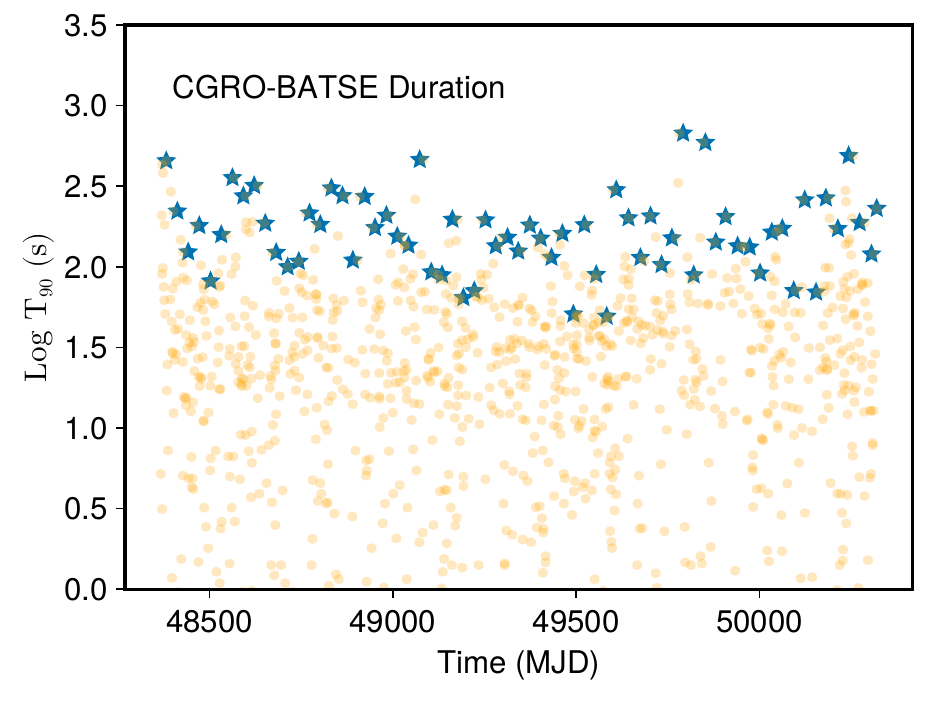} &
\includegraphics[width=0.6\columnwidth]{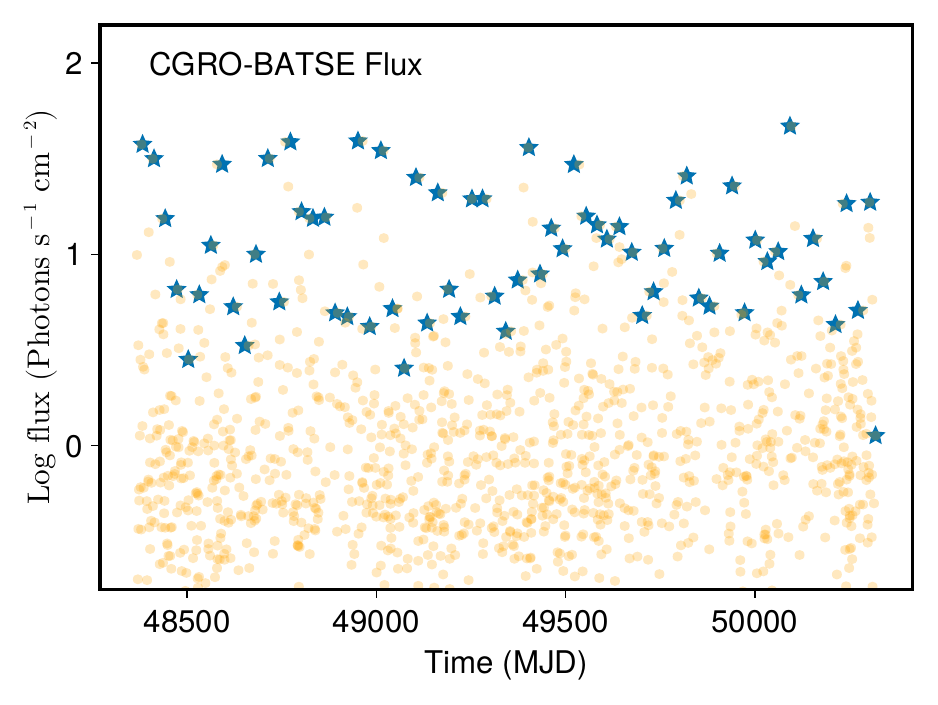} & \includegraphics[width=0.6\columnwidth]{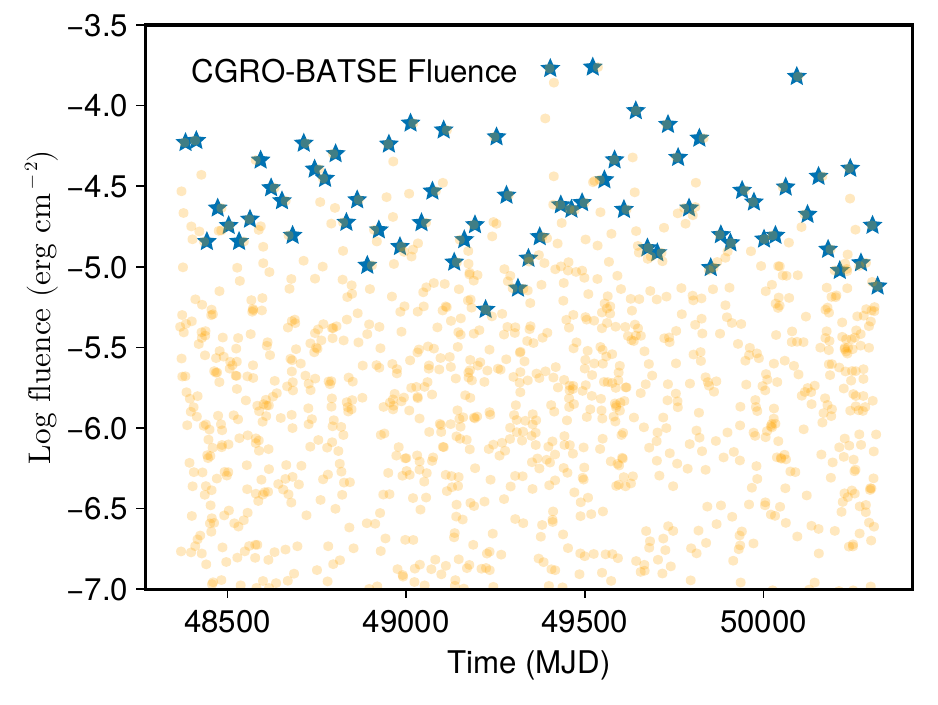} \\
\includegraphics[width=0.6\columnwidth]{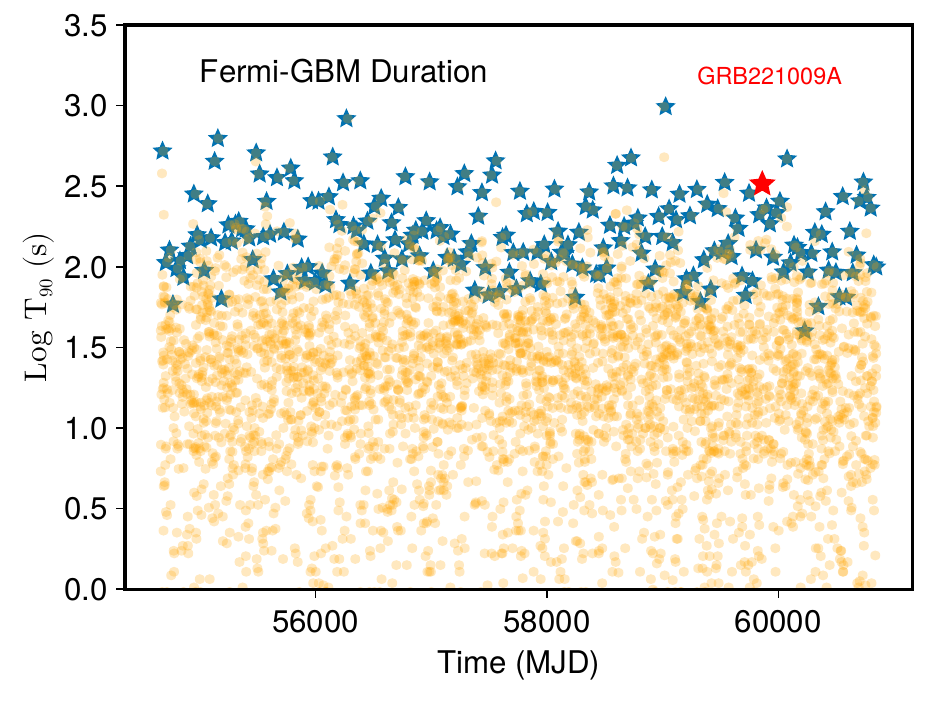} &
\includegraphics[width=0.6\columnwidth]{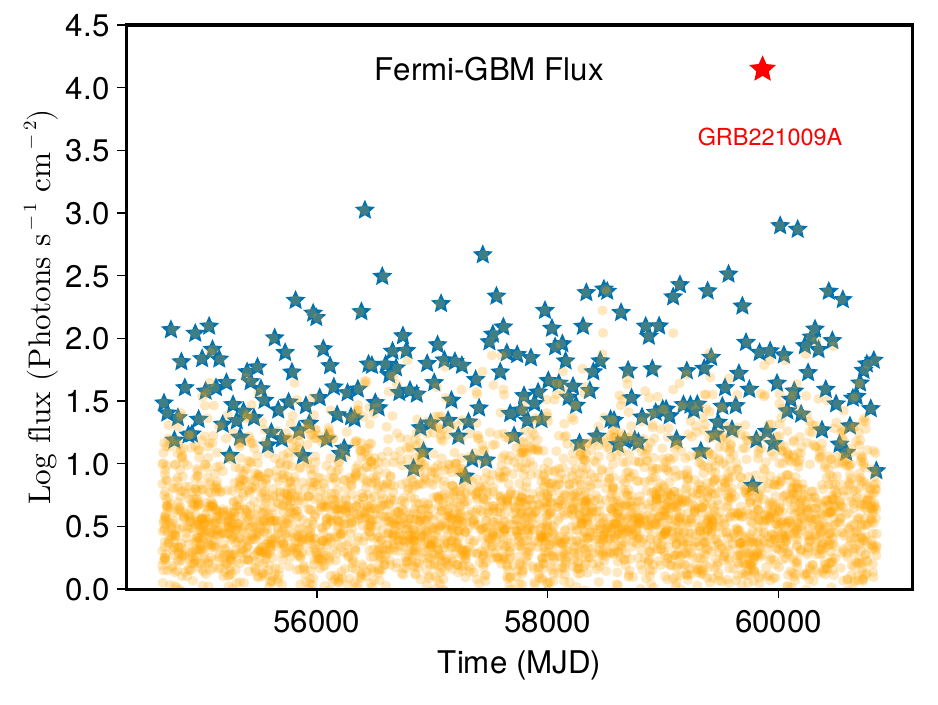} &
\includegraphics[width=0.6\columnwidth]{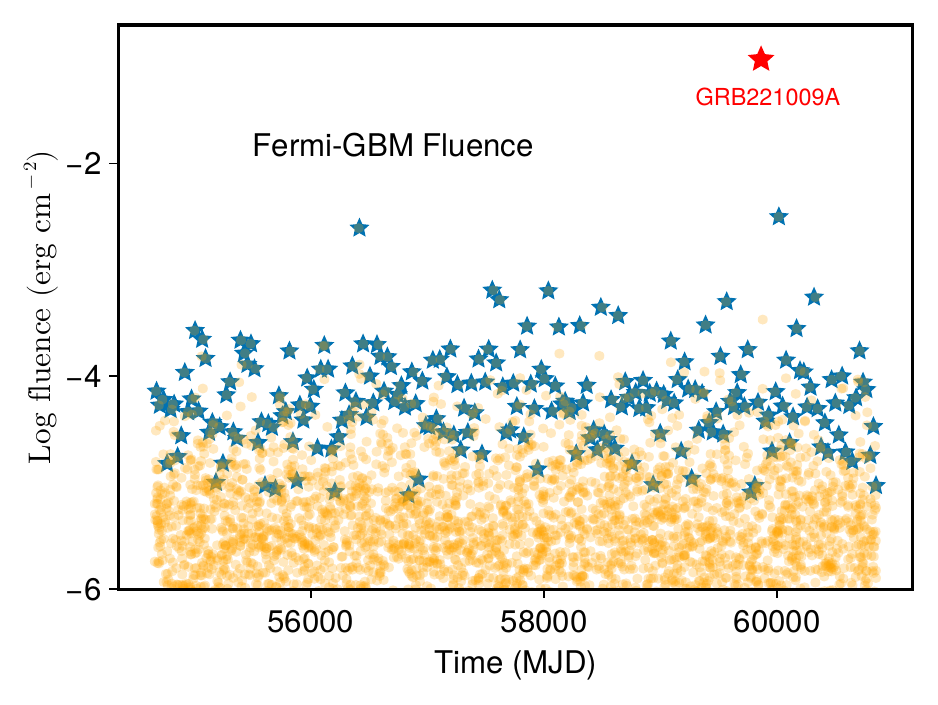}
\end{tabular}
\caption{Extreme values (blue stars) vs the whole population of GRBs (yellow points) for data from the BATSE catalogue (top) and GBM catalogue (bottom). From the left to the right, we show the duration, the peak flux and the fluence. The fluence and peak flux of \object{GRB\,221009A} stands out more than two orders of magnitude above the bulk of the GRBs detected by Fermi, while for the duration the event is not peculiar \citep[see, e.g., discussion in ][]{Burnsetal2023}. A red star indicates the actual location of \object{GRB\,221009A} in each plot.}
\label{fig:gev}
\end{figure*}

We show here (Fig.\,\ref{fig:gev}) the extreme values for fluences, peak fluxes, and durations for GRB data reported in the CGRO-BATSE and Fermi-GBM catalogues in bins of 30\,days.

\section{Corner plots and fitted values}

\begin{figure*}[hb]
\centering
\begin{tabular}{ccc}
\includegraphics[width=0.6\columnwidth]{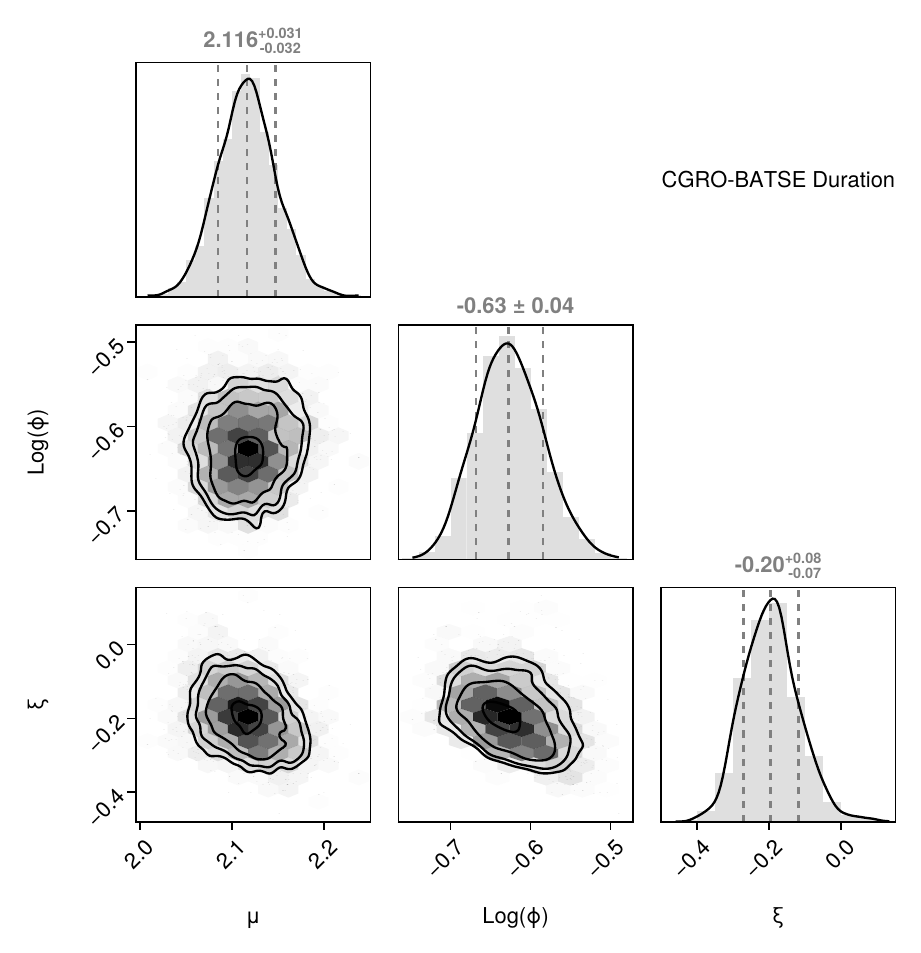} &
\includegraphics[width=0.6\columnwidth]{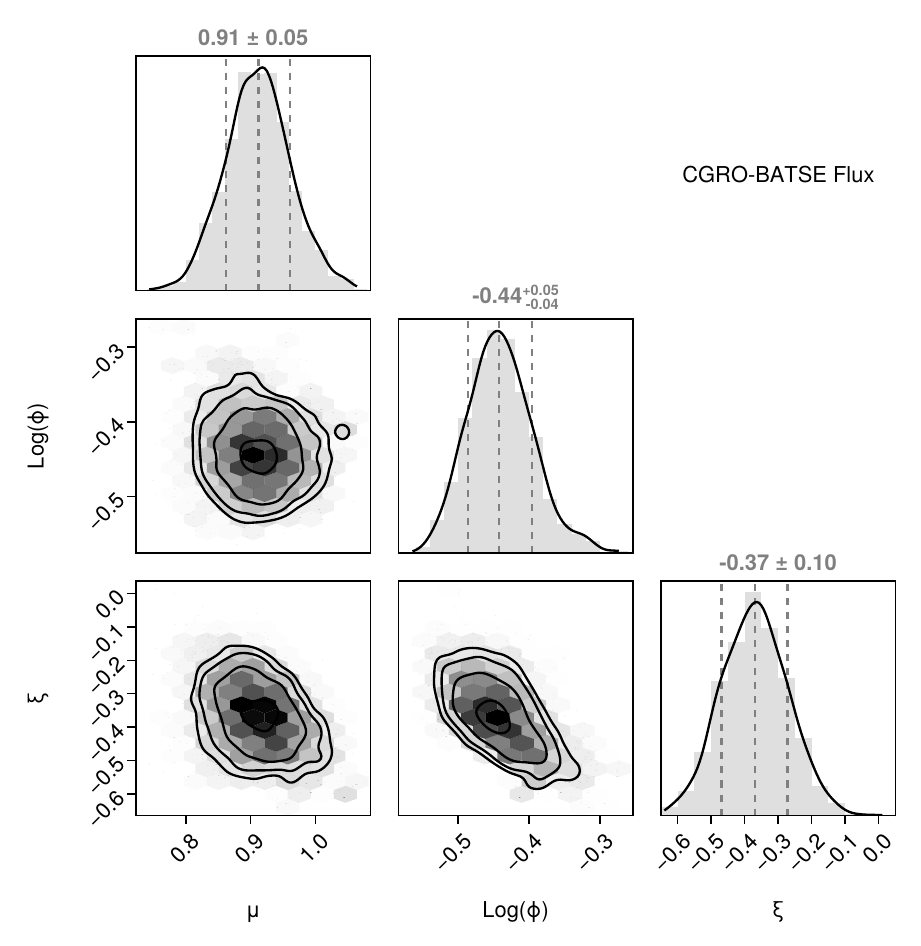} &
\includegraphics[width=0.6\columnwidth]{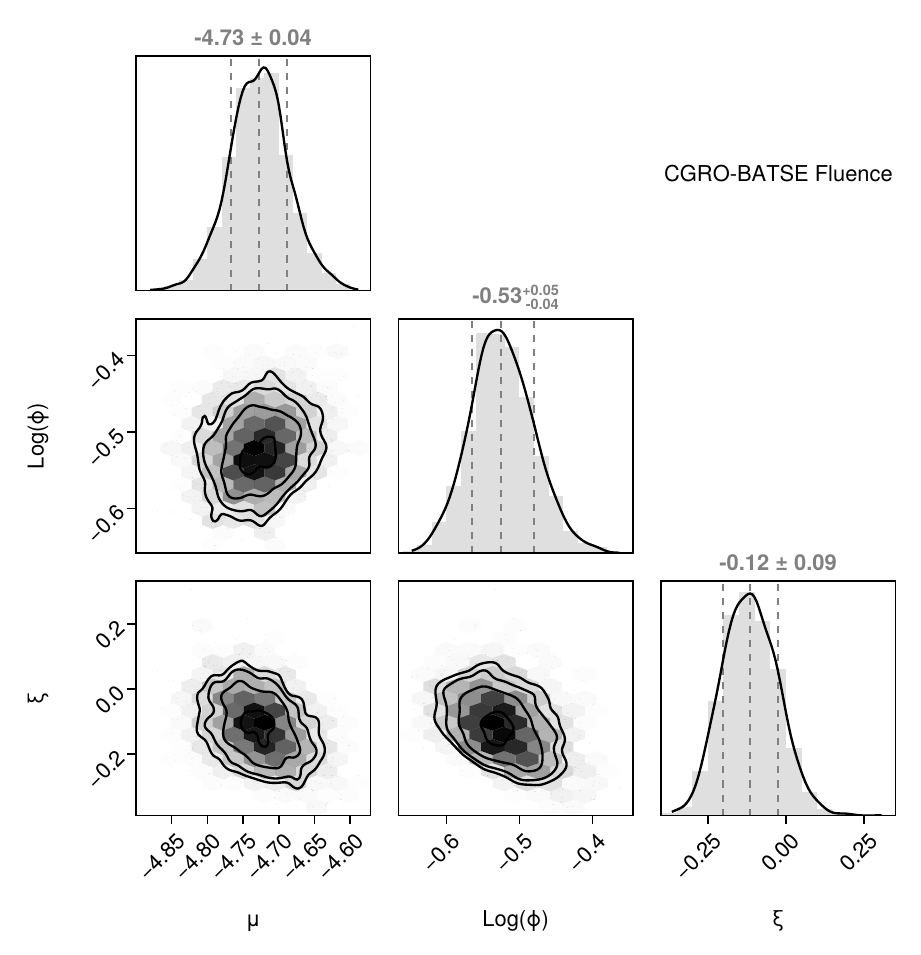} \\
\includegraphics[width=0.6\columnwidth]{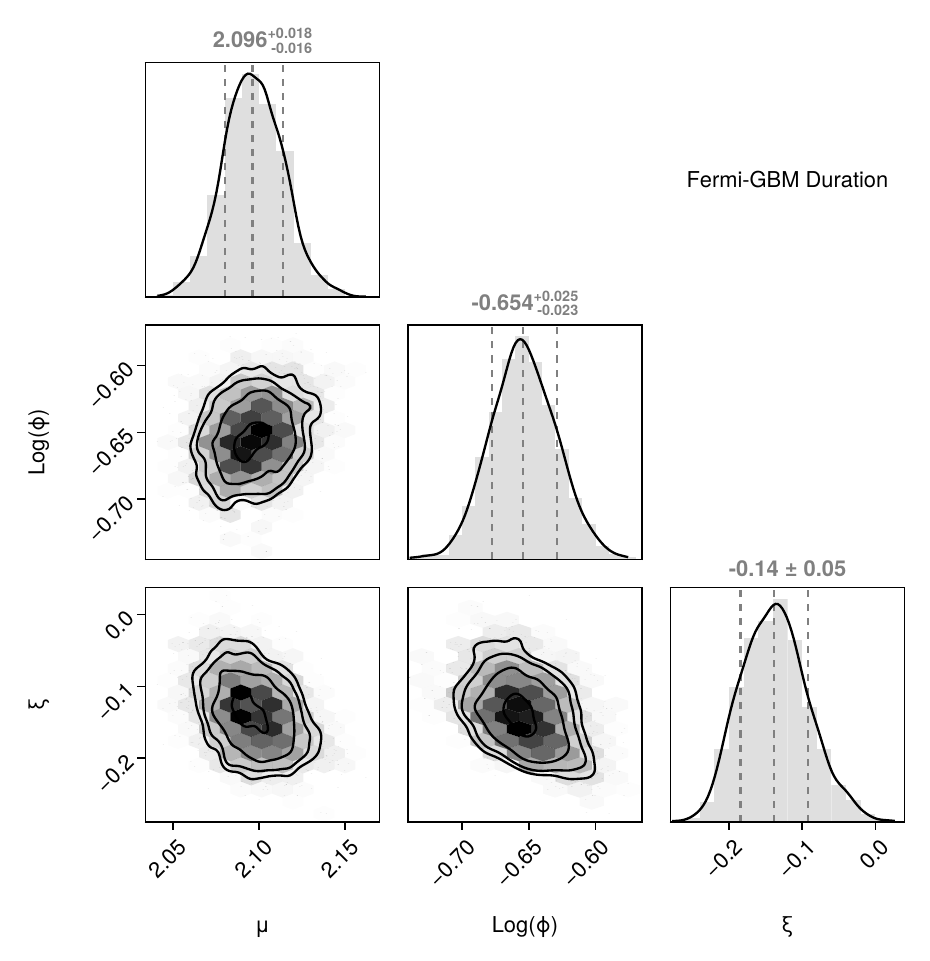} &
\includegraphics[width=0.6\columnwidth]{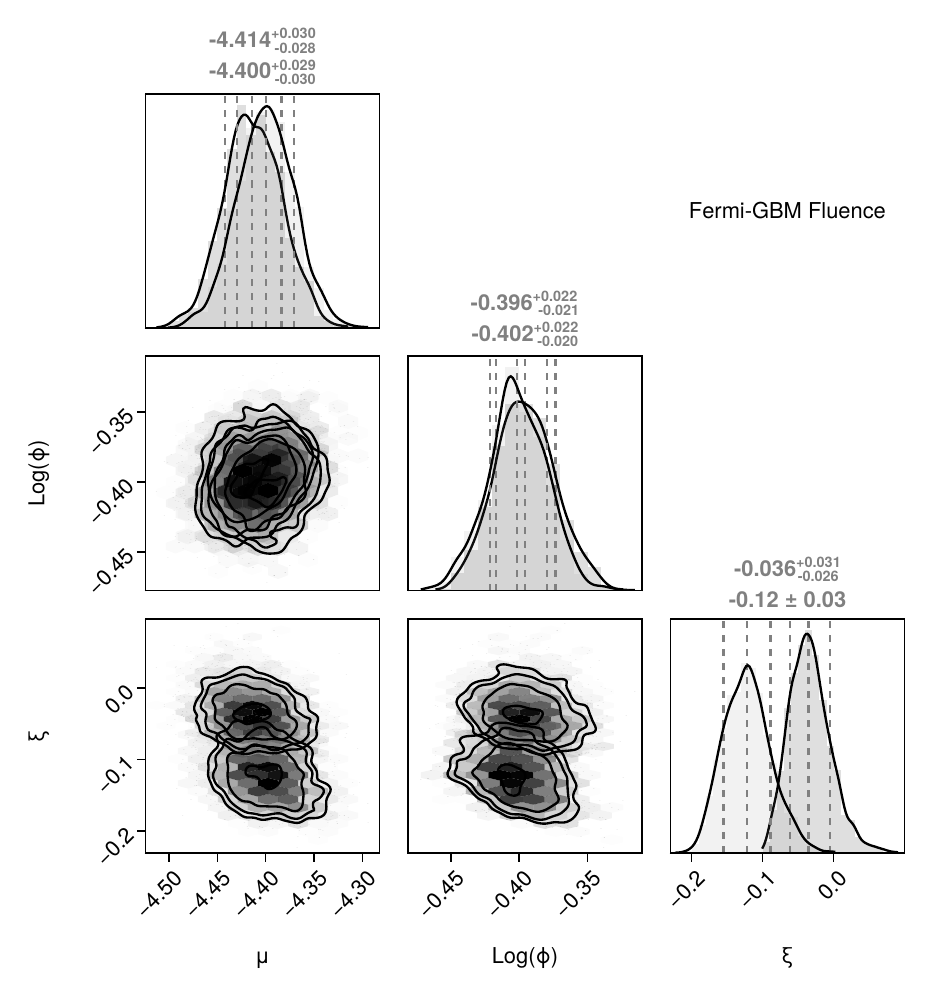} &
\includegraphics[width=0.6\columnwidth]{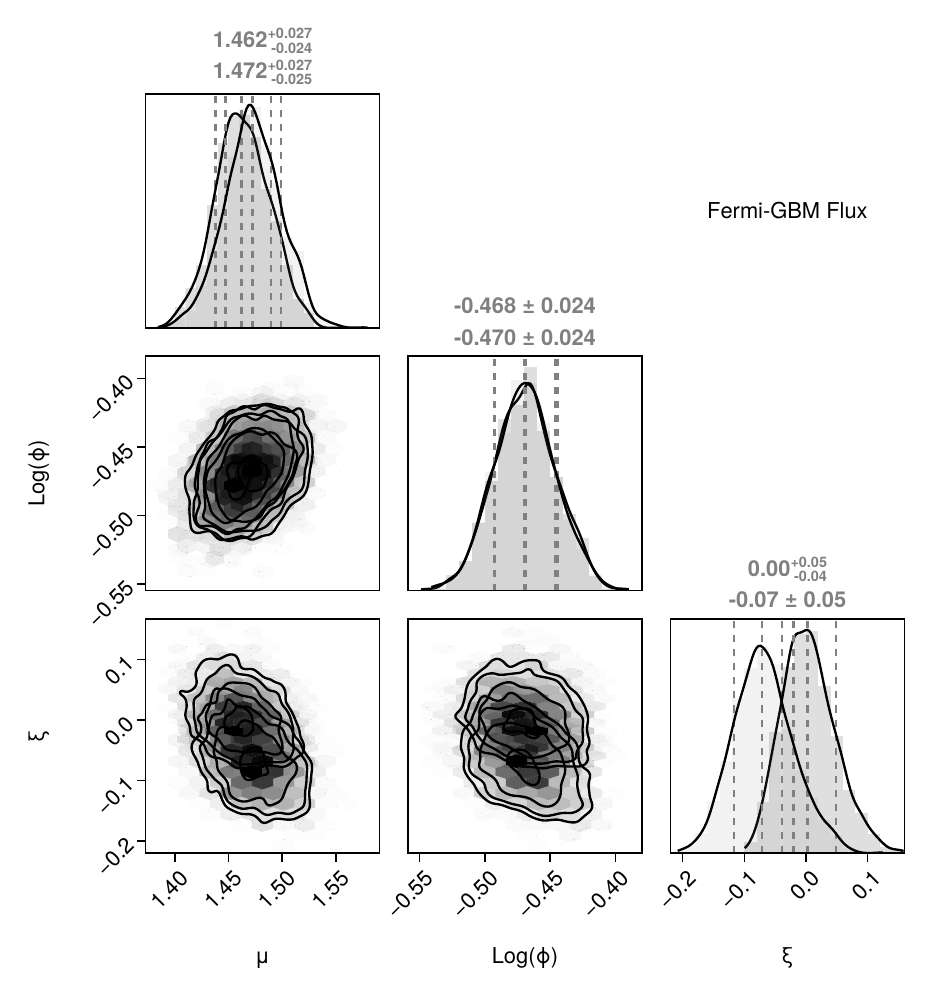} \\
\end{tabular}
\caption{Corner plots for the posterior distributions of the GEV parameters for data from the BATSE catalogue (top) and GBM catalogue (bottom). From the left to the right we have results for the duration, the peak flux and the fluence. The sampled parameters are the {\it \emph{mean}}, $\mu$, the {\it \emph{scale}}, $\phi$, and the {\it \emph{shape}}, $\xi$ of the GEV distribution. For the GBM dataset we carried out the analysis with or without the extreme \object{GRB\,221009A}. The removal of this event affects only the GEV distributions for the fluence and peak flux.}
\label{fig:corners}
\end{figure*}

We show here (Fig.\,\ref{fig:corners}) the "corner plots" derived by the Bayesian analysis of extreme values for fluences, peak fluxes, and durations for GRB data reported in the CGRO-BATSE and Fermi-GBM catalogues. Summary statistics for the fitted parameters are shown in Table\,\ref{tab:fitres}. Errors on the credible intervals are reported at 1$\sigma$ level.

\begin{table}[h]
\caption{GEV fitted parameters and uncertainties}
\label{tab:fitres}
\begin{small}
\begin{tabular}{c c c c c}
\hline
\hline
\textbf{Fluence}   & CGRO-BATSE & Fermi-GBM & Fermi-GBM  \\
   &            &           &   (without the BOAT) \\
\hline
$\mu$ & $-4.73 \pm 0.04$ & $-4.41 \pm 0.03$ & $-4.40 \pm 0.03$ \\
$\log \phi$ & $-0.53^{+0.05}_{-0.04}$ & $-0.40 \pm 0.02$ & $-0.40 \pm 0.02$ \\
$\xi$ & $-0.12 \pm 0.09$ & $-0.04 \pm 0.03$ & $-0.12 \pm 0.03$ \\
\hline
\textbf{Peak flux}   & CGRO-BATSE & Fermi-GBM & Fermi-GBM  \\
   &            &           &   (without the BOAT) \\
\hline
$\mu$ & $0.91 \pm 0.05$ & $1.46^{+0.03}_{-0.02}$ & $1.47 \pm 0.03$ \\
$\log \phi$ & $-0.44^{+0.05}_{-0.04}$ & $-0.47 \pm 0.02$ & $-0.47 \pm 0.02$ \\
$\xi$ & $-0.37 \pm 0.10$ & $0.00^{+0.05}_{-0.04}$ & $-0.07 \pm 0.05$ \\
\hline
\textbf{Duration}   & CGRO-BATSE & Fermi-GBM & Fermi-GBM  \\
   &            &           &   (without the BOAT) \\
\hline
$\mu$ & $2.12 \pm 0.03$ & $2.10 \pm 0.02$ &  \\
$\log \phi$ & $-0.63 \pm 0.04$ & $-0.65^{+0.03}_{-0.02}$ &  \\
$\xi$ & $-0.20^{+0.08}_{-0.07}$ & $-0.14 \pm 0.04$ &  \\
\hline
\end{tabular}
\end{small}
\tablefoot{The $1\sigma$ interval and the average of the fitted parameters of the derived GEV distributions. We report results for the extreme values computed from the CGRO-BATSE catalogue and from the full Fermi-GBM catalogue. For comparison, we also show results obtained having removed \object{GRB\,221009A} from the catalogue for the fluence and peak flux.}
\end{table}

\end{appendix}

\end{document}